\documentclass[preprint]{elsarticle}

\usepackage{hyperref}
\hypersetup{colorlinks=true}

\journal{arXiv.org cs.HC}
\biboptions{authoryear}

\usepackage{tabu}     
\usepackage{booktabs} 
\usepackage{gensymb}  

\usepackage{fancyhdr} 
\usepackage{lastpage} 
\usepackage{xpatch}
\xpatchcmd{\pprintMaketitle}{\hrule}{}{}{} 
\xpatchcmd{\pprintMaketitle}{\hrule}{}{}{} 

\begin{document}

\begin{frontmatter}

\title{User Preferences of Spatio-Temporal Referencing Approaches For Immersive 3D Radar Charts}

\author[vrxaraddress,ivisaddress]{Nico Reski\corref{correspondingauthor}}
\ead{nico.reski@liu.se}

\author[vrxaraddress]{Aris Alissandrakis\corref{correspondingauthor}}
\cortext[correspondingauthor]{Corresponding author}
\ead{aris.alissandrakis@lnu.se}

\author[isovisaddress,ivisaddress]{Andreas Kerren}
\ead{andreas.kerren@liu.se}

\address[vrxaraddress]{VRxAR Labs, Department of Computer Science and Media Technology, Linnaeus University, V\"axj\"o, Sweden}
\address[isovisaddress]{ISOVIS, Department of Computer Science and Media Technology, Linnaeus University, V\"axj\"o, Sweden}
\address[ivisaddress]{iVis, Department of Science and Technology, Link\"oping University, Norrk\"oping, Sweden}

\begin{abstract}
The use of head-mounted display technologies for virtual reality experiences is inherently single-user-centred, allowing for the visual immersion of its user in the computer-generated environment.
This isolates them from their physical surroundings, effectively preventing external visual information cues, such as the pointing and referral to an artifact by another user.
However, such input is important and desired in collaborative scenarios when exploring and analyzing data in virtual environments together with a peer.
In this article, we investigate different designs for making spatio-temporal references, i.e., visually highlighting virtual data artifacts, within the context of Collaborative Immersive Analytics.
The ability to make references to data is foundational for collaboration, affecting aspects such as awareness, attention, and common ground.
Based on three design options, we implemented a variety of approaches to make spatial and temporal references in an immersive virtual reality environment that featured abstract visualization of spatio-temporal data as 3D Radar Charts.
We conducted a user study (n=12) to empirically evaluate aspects such as aesthetic appeal, legibility, and general user preference.
The results indicate a unified favour for the presented \textit{location} approach as a spatial reference while revealing trends towards a preference of mixed temporal reference approaches dependent on the task configuration: \textit{pointer} for elementary, and \textit{outline} for synoptic references.
Based on immersive data visualization complexity as well as task reference configuration, we argue that it can be beneficial to explore multiple reference approaches as collaborative information cues, as opposed to following a rather uniform user interface design. 
\end{abstract}

\begin{keyword}
awareness\sep
collaborative immersive analytics\sep
computer-supported cooperative work\sep
empirical study\sep
virtual reality\sep
visual information cues\sep
3D radar chart
\end{keyword}

\end{frontmatter}


\pagestyle{fancy} 
\lhead{\footnotesize \emph{Preprint submitted to arXiv.org cs.HC}} \chead{}\rhead{\footnotesize \emph{\today}}
\lfoot{}\cfoot{\footnotesize \thepage\ of \pageref*{LastPage}}\rfoot{}
\renewcommand{\headrulewidth}{0pt}

\section{Introduction}\label{sec:introduction}

Utilizing immersive display and interaction technologies for the purpose of data exploration, analytical reasoning, and decision making, i.e., Immersive Analytics (IA), has become an increasingly intriguing research domain as relevant hardware and software technologies advance in affordability, accessibility, and usability \citep{dwyer2018iai,skarbez2019iat}.
\citet{fonnet2021soi} found that the amount of publications presenting systems that feature three-dimensional (3D) graphics, stereoscopic vision capabilities, and head-tracking support within the context of IA has noticeably increased since 2012.
Data interpretation and subsequent analytical reasoning often rely on collaboration, i.e., multiple analysts combine knowledge and expertise to discuss their findings and to make decisions -- a process that is much desired by the community \citep{hackathorn2016iab,heer2008dcf,isenberg2011cvd}.
However, the by default single user-centred characteristics of many immersive interfaces are often in conflict with such anticipated collaboration, requiring careful design considerations to support effective collaboration of multiple users \citep{billinghurst2018cia,ens2019rct,skarbez2019iat}.
Research published within the domain of Computer-Supported Cooperative Work (CSCW) shows that the understanding, support, and evaluation of collaboration when using information and communication technologies is a complex task that demands the consideration of many different aspects and dimensions, for instance as elaborated by various descriptive models and frameworks \citep{gutwin2002dfw,isenberg2012ccv,johansen1988gcs,lee2015ftm,neumayr2018ddf,tang2006cco}.
\citet{churchill1998cve} already described in 1998 important characteristics that should be addressed when designing for collaboration in Virtual Environments (VEs) -- a topic still relevant today.
Among others, collaborative VEs should strive to enable users to share their context and make them aware of each other, as well as allow them to discuss their respective findings \citep{churchill1998cve}.
As an inherently multi-disciplinary research area, there is a lot of potential to utilize existing knowledge, guidelines, and recommendations when designing Collaborative Immersive Analytics (CIA) experiences \citep{dwyer2018iai,skarbez2019iat}.
For instance, concepts such as Awareness \citep{gutwin2002dfw,heer2008dcf}, Communication, Cooperation, Co-ordination \citep{andriessen2003gp}, Reference, and Deixis \citep{heer2008dcf} provide foundational concepts that can be utilized to implement anticipated collaborative features to support data analysts accordingly.

Support for collaboration within the context of IA has been deemed integral for its success and establishment of valuable data analysis tools, with several research challenges requiring further exploration \citep{ens2021gci}.
A promising direction to enable collaboration in VEs, and thus support various collaborative information cues, is the use of avatars, i.e., a virtual representation of the other user(s) in the VE, either co-located or connected remotely \citep{steed2015cii,xi2018sef}.
The visual design of such avatars has been investigated in a multitude of studies, for instance, to determine differences between realistic and abstract avatar representations \citep{sun2019nsi}, to explore effects of nonverbal expression using highly expressive avatars \citep{wu2021uaf}, or to determine how avatar appearances influence aspects of communication and interaction \citep{heidicker2017ioa}.
Such collaboration and the use of avatars often imply that all collaborators are part of the same immersive VE, i.e., using homogeneous device types and technologies that enable them to share their 3D information space.
However, such an approach is not always feasible, for instance when using heterogeneous display and interaction technologies to collaborate on the same data across different applications.
Particularly within the context of CIA, the mixture and integration of data analysis tools that are based on different technologies are highly anticipated, in order to build workflows that synergize with interactive Information Visualization (InfoVis) and Visual Analytics (VA) tools \citep{ens2021gci,isenberg2014aic,wang2019avo}.

With this in mind, we are interested in the investigation of design approaches for making visual references in immersive data visualizations, assisting the immersed analyst to focus their attention towards a point of reference as indicated through a non-immersed collaborator.
Based on different design options, we implemented various approaches to make spatial and temporal references inside a Virtual Reality (VR) environment that enables an analyst to explore a multivariate dataset using a head-mounted display (HMD).
We envision that such visual references facilitate natural communication and coordination with the immersed user, especially in scenarios where analysts use heterogeneous display and interaction technologies (hybrid) in order to have individual perspectives and roles during the data analysis (asymmetric) \citep{ens2021gci,isenberg2014aic,wang2019avo}.
As the HMD user is visually isolated from their real-world surroundings, common means of referencing through a collaborator (e.g., pointing to an artifact) are no longer available, even in co-located work spaces.
Consequently, there is a need for adequate means of referencing as a sufficient replacement.
Our research focus is therefore concerned with the design of visual references as nonverbal information cues, particularly within the context of immersive spatio-temporal data visualization.
This allows us to contribute to the emerging field of CIA as follows:

\begin{itemize}
    \setlength\itemsep{0em}
    \item We describe generalized options for the design of visual references in order to make spatial and temporal references in immersive VR data visualizations.
    \item We report on the results of an empirical evaluation that used subjective methods to examine several implemented visual referencing approaches in regard to aesthetics, legibility, and general user preference.
    \item We discuss the results and reflect on the implemented referencing approaches, providing directions that can guide the design of similar collaborative information cues.
\end{itemize}

The structure of the article is as follows: Section~\ref{sec:relatedwork} provides an overview of relevant CSCW terminology, describes existing work of design approaches to integrate collaborative information cues in VR applications, and provides additional motivation for the support of referencing within the context of CIA.
We describe the reference design options and subsequent implementation for the various spatio-temporal reference approaches in Section~\ref{sec:spatiotemporalreferences}.
We conducted an empirical evaluation to assess aesthetics, legibility, and general user preference for all presented reference approaches.
Section~\ref{sec:methodology} describes the applied evaluation methodology, providing details on the study design and measures.
Based on the results as presented in Section~\ref{sec:results}, we discuss the gathered data thereafter in Section~\ref{sec:discussion}, contributing with reflections on the visual reference design for their consideration in similar applications.
Finally, we conclude by summarizing our work and providing some directions for future work in Section~\ref{sec:conclusion}.

\section{Related Work}\label{sec:relatedwork}

\subsection{Understanding Relevant CSCW Terminology}\label{sec:cscwterminology}

Rich CSCW terminology exists to describe various aspects and dimensions of collaboration, especially for scenarios of synchronous collaboration, i.e., multiple users working on a subject matter at the same time.
Arguably, it can be difficult to differentiate terms such as awareness, attention, common ground, focus, referencing, and so forth, especially as some might be used ambiguously \citep{schmidt2002tpw}.
We adopt CSCW terminology as follows within the scope of our investigation.

\citet{schmidt2002tpw} elaborates on the many facets of \textit{Awareness}, describing it as a collaborator's ability to align and integrate their own actions with those of others, without interruption and major efforts, in an organic way.
\citet{heer2008dcf} expand on that by stating that \textit{Awareness} should allow for the assessment of work and task completion, enabling each user to make decisions on where to allocate their next efforts.
Closely aligned with awareness are also the concepts of \textit{Common Ground}, describing the state of the collaborators' shared understanding as the foundation to enable communication among them accordingly \citep{clark1991gic}, and \textit{Grounding} as the process of achieving that state \citep{heer2008dcf}.
\citet{andriessen2003gp} provides a differentiation between \textit{Communication}, \textit{Co-operation}, and \textit{Co-ordination} as general group processes.
They categorize \textit{Communication} as an interpersonal exchange process, utilizing tools for the exchange of (verbal and nonverbal) signals \citep{andriessen2003gp}.
\textit{Co-operation} is described as the task-oriented process of collaborators actually working together, making decisions, and co-manipulating artifacts.
Also categorized as a task-oriented process is the concept of \textit{Co-ordination}, enabling the collaborators to adjust their individual and group efforts to solve a given task.
\textit{Attention} \citep{kristoffersen1999mpt,shneidermann2017dtu} and \textit{Focus} \citep{schmidt2002tpw,snowdon2001cve} within the context of collaboration are often seen ambiguously, generally referring to a user's visual alignment to a specific area or point of interest in the shared workspace. 
When working together, a frequently used communication action is to make a \textit{Reference}, for instance in order to indicate a specific object, area, person, or time, often through a mixture of verbal (e.g., talk) and nonverbal (e.g., point)  signals \citep{heer2008dcf}.

Dissecting collaborative work is a complex endeavor, as many states and actions are inherently interconnected and dependent on each other.
To describe just one example, the ability to make a reference as means of communication allows a collaborator to focus and pay attention to respectively referred artifacts, impacting aspects of each individual's awareness as well as the team's coordination and common ground.
Such aspects are essential for the team's joint exploration, interpretation, and discussion of data.
It is apparent that careful design considerations are required when creating systems and environments that aim to support such collaborative work.

\subsection{Collaborative Information Cues in Virtual Reality}\label{sec:civsinvr}

Providing means that aid users in their communication is vital for their collaboration, as emphasized in Section~\ref{sec:cscwterminology}.
The ability to make visual references is arguably even more important in hybrid technology scenarios where not every collaborator is immersed in VR, requiring features to bridge such shortcomings.
Some interesting work has been conducted to provide collaborative information cues in immersive VR environments.

Considering an asymmetric user role setup, \citet{peter2018vrg} presented a tool for a non-immersed \textit{VR-Guide}, providing features to support the guidance of an immersed VR user.
Their implemented proof-of-concept prototype included three visual approaches as non-verbal information cues, aiming to catch the VR user's attention and thus guide them towards an artifact as selected through the non-immersed guide. 
First, they implemented an outline effect, visually highlighting the border of the selected artifact independent of its occlusion, i.e., the outline is visible even with other objects in the VE between the user and the targeted artifact, leaving it otherwise hidden from the user's view. 
Second, they implemented a realistic-looking light beam technique, similar to a spotlight, allowing the highlighted artifact to be visually distinguished from others. 
And third, a virtual drone was implemented using a laser beam to point to the selected artifact. 
An interesting aspect of their tool is to allow the VR-Guide to customize these signals, for instance by adjusting the color or thickness of the outline effect, or the size and intensity of the spotlight. 
Their evaluation revealed trends towards a higher acceptance of the outline technique compared to the light beam one, with participants interestingly expressing a desire for a combination of outline and virtual drone techniques. 

A visually different approach was presented by \citet{sugiura2018aac}, implementing a virtual hand in a pointing hand posture that literally points to the selected artifact.
Their approach was partially inspired by previous work of \citet{stafford2006iog} but adopted to support a setup that involved the non-immersed user operating an interactive tabletop application, allowing touch interaction to select individual objects from a top-down view that are referred to accordingly in VR \citep{sugiura2018aac}.
The results of their preliminary user study suggest the overall usability of their prototype but lack a formal evaluation of the effectiveness of the implemented guiding technique. 

\citet{welsford2020aib} evaluated a prototype that allowed interaction between an HMD user and a non-HMD one that operates a large-scale immersive display.
A laser pointer technique enabled the non-HMD user, in a more ``spectator'' role, to make a visual reference that would create a virtual marker in VR accordingly for the HMD user. 
The results of their evaluation indicate that the users could communicate effectively using the provided features. 

\citet{lacoche2017caf} investigated different visual approaches to raise awareness towards other users in a co-located collaborative VR environment, aiming to prevent physical collisions.
First, an extended grid representation visualizes the other user's position through a grid-shaped cylinder, allowing a VR user to avoid navigating to the same position, while still being able to ``look beyond'' (preventing occlusion issues). 
Second, another user was indicated through means of a ``Ghost Avatar'', displaying some features of that user's position and orientation, e.g., a semi-transparent model of their HMD and hand controllers. 
Third, a safe navigation floor utilizes a heat map-inspired approach to visualize on the virtual floor where it is safe to go and where not, avoiding collision with the other user as well as a movement beyond the physical boundaries of the VR area. 
Under the introduction of a fourth approach (separated tracked spaces) that limited each user's designated VR area as a typical bounding box grid, an evaluation was conducted to investigate the effectiveness and user preference for the different approaches. 
Their results point towards better performance of the extended grid and ghost avatar compared to the safe navigation floor approach. 

\citet{chen2018idm} investigated different modalities (visual, auditory, vibrotactile) to make directional information cues during a multi-task scenario in VR.
Based on task completion time and accuracy measurements, the results of their study indicate a favor towards direction cues based on visual and vibrotactile stimuli over auditory ones. 

\citet{casarin2018uau} described a developed toolkit that allows synchronous interaction in VEs through multiple users.
To avoid simultaneous manipulation of the same artifact, the authors implemented an abstract interaction filter to visually indicate whether or not an artifact is available for interaction. 
Three different states (with applied color coding) provided visual feedback in real-time: An artifact is available for interaction (original color), hovered (green color), or currently being manipulated (orange color). 
Their toolkit was validated based on the results of an evaluation that featured a collaborative authoring task, but further investigations are necessary in order to address individual aspects of their toolkit, such as for instance the design of the collaborative information cues.

\subsection{Motivation to Support Referencing in CIA}

\begin{figure}
    \centering
    \includegraphics[width=0.9\columnwidth]{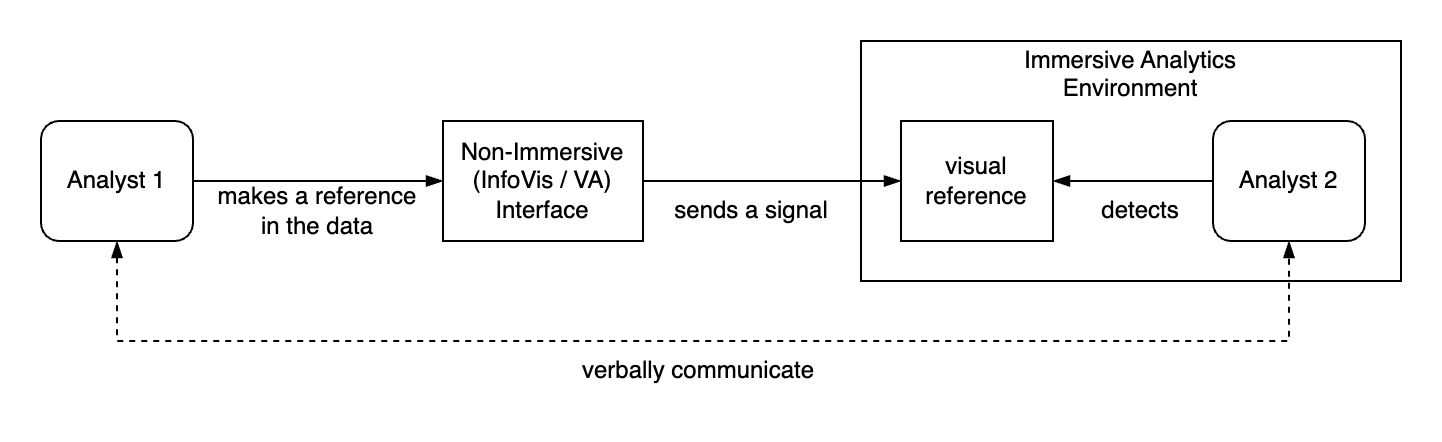}
    \caption{Context and scenario of the research focus.}
    \label{fig:researchfocussetup}
\end{figure}

A recent survey on IA-related research reveals a lack of work in regard to collaborative systems \citep{fonnet2021soi}.
Considering the importance of collaboration within that context \citep{ens2021gci}, further investigations are needed to provide design recommendations for CIA experiences.
The work presented throughout Sections~\ref{sec:cscwterminology} and~\ref{sec:civsinvr} provides important directions and guidance to further examine similar matters.
In particular, we are motivated to investigate the design of collaborative information cues within the context of CIA as follows.
As a starting point, we assume that an analyst is utilizing HMD technologies to explore data in immersive VR.
To support collaborative exploration and interpretation of data, the VE needs to provide features that aid the VR user to follow their collaborator's signals, e.g., \textit{nonverbal} references.
Within the scope of this work, we focus on \textit{visual} references, as we anticipate active verbal communication between the analysts during their synchronous collaboration. 
Such references aim to facilitate the VR user's ability to focus their attention toward an indicated artifact in the VE, allowing the subsequent establishment of common ground between the collaborators to further discuss and interpret the data.
As best to our knowledge, similar referencing and guiding approaches are yet to be investigated in regard to abstract data visualization.
\autoref{fig:researchfocussetup} illustrates the overall context and scenario of the described research focus.

\section{Spatio-Temporal Referencing in VR}\label{sec:spatiotemporalreferences}

\begin{figure}
    \centering
    \includegraphics[width=.9\columnwidth]{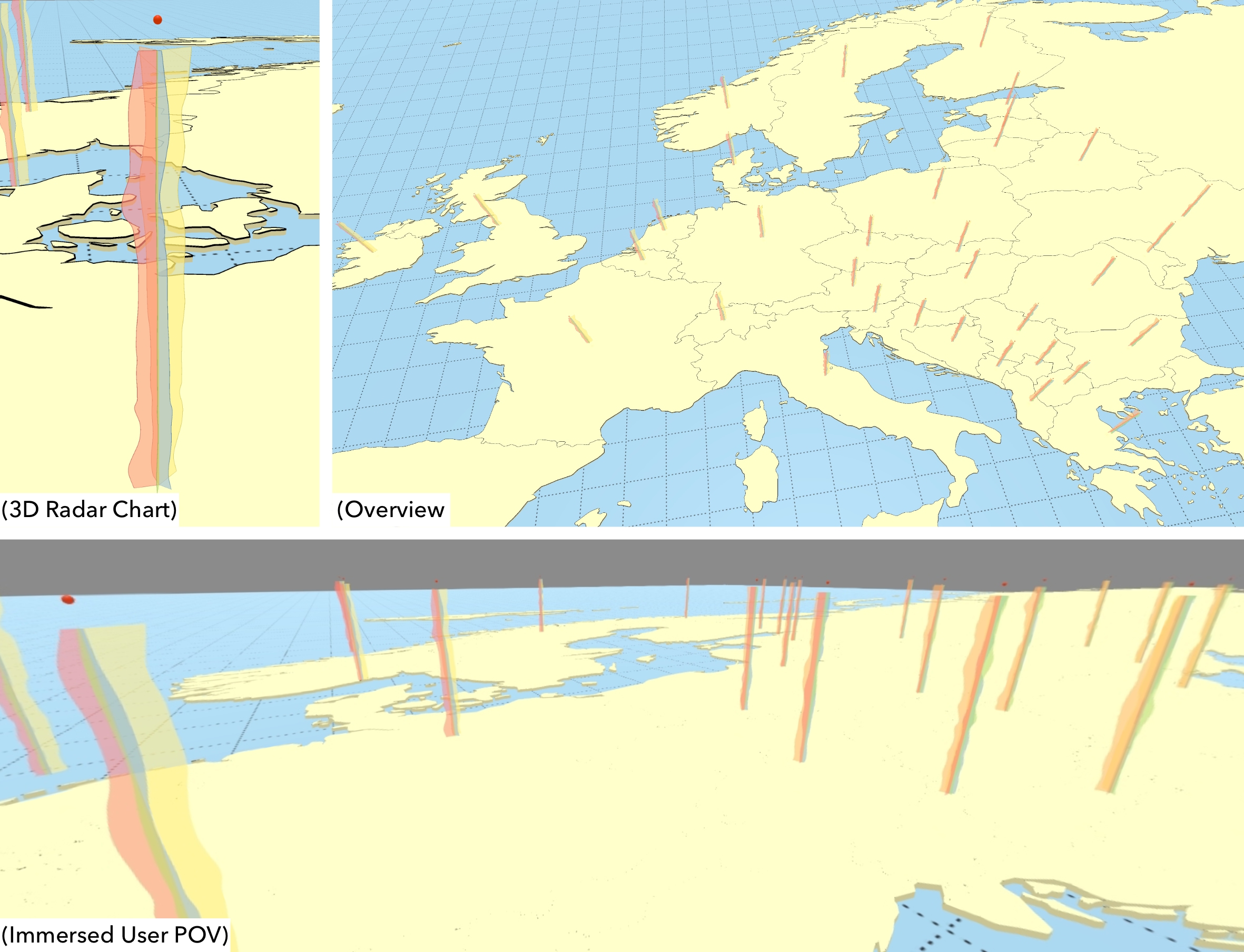}
    \caption{Spatio-Temporal Data Visualization in VR, as adopted from \citet{reski2020eot}.
    \textbf{Top Left}: 3D Radar Chart visualization.
    \textbf{Top Right}: Excerpt of the VE, taken from an angled-top down perspective to provide an overview impression.
    \textbf{Bottom}: VE from the immersed user's point of view (POV).
    \textbf{General description}:
    Individual 3D Radar Charts are spatially arranged to represent individual European countries, which are visually differentiated on the virtual floor as extruded polygons.
    Each 3D Radar Chart here features five data variables as color-coded semi-transparent frequency polygons, each including 150 consecutive time events as a time series.
    The presented data has been generated artificially for illustration purposes.}
    \label{fig:overviewimmersivedatavis}
\end{figure}

Visualizing spatio-temporal data using immersive technologies is a comparatively common use case \citep{fonnet2021soi}.
It is arguably easy to conceive the utilization of the additional (third) dimension for the mapping of data variables.
For instance, aspects of a multivariate dataset in regard to spatial dimensions can be placed in respective relation to each other, whereas common two-dimensional (2D) visualization techniques have the potential to be expanded into the 3D space, e.g., displaying time-oriented data \citep[Chapter~7]{aigner2011vot}.
Several toolkits have been recently presented, aiming to assist the practical implementation of (abstract) 3D visualizations within the context of IA \citep{butcher2019vaf,cordeil2019iai,sicat2019dxr}.
Within the scope of our investigation, we adopt the approach of \textit{3D Radar Charts} as immersive time visualization as presented by \citet{reski2020eot}.
In this approach, different time-series data variables are placed as 2D frequency polygons in a radial arrangement around a time axis in 3D, allowing for the visual exploration of the data dimensions \citep{reski2020eot}.
Individual values of these displayed data variables can be connected to each other, representing the more traditional radar chart pattern \citep{kolence1973sup}, enabling further interpretation and interaction with the time-series data \citep{reski2020eot}.
In practice, the virtual 3D space can be populated through the placement of multiple such 3D Radar Chart visualizations, each representing a spatial dimension in a dataset (e.g., country, municipality, city, or GPS coordinates).
This allows an immersed analyst to explore data in regard to both spatial and temporal dimensions.
\autoref{fig:overviewimmersivedatavis} illustrates the setup of such a virtual analysis environment, serving as the foundation for the practical implementation of our spatial and temporal reference design approaches according to the described design options and task types presented throughout the following sections.

\subsection{Design Options}\label{sec:designoptions}

To provide a reference that catches the user's attention and thus guides them toward an artifact, a signal is required that allows to be distinguished from the conventional environment. 
Within the scope of our investigation, we focus on \textit{visual} sensory input, i.e., the manipulation of the immersive data visualization through means that allow its user to visually perceive references by looking around.\footnote{At this stage, we do not consider other sensory input, such as for instance from auditory or haptic interfaces.}
We defined three design options to guide the creation of visual references as collaborative information cues:
\textit{Modify Artifact}, \textit{Add Artifact}, and \textit{Modify Environment}.
These three design options are held purposefully generic and low-level to allow application across many different scenarios and use cases (see \autoref{tab:designoptions}).

\begin{table}
  \caption{Design Options -- Categorization of the implemented reference design approaches according to the presented work (see Sections~\ref{sec:spatialreferencedesign} and~\ref{sec:temporalreferencedesign}) as well as the related work (see Section~\ref{sec:civsinvr}).}
  \label{tab:designoptions}
  \scriptsize
	\centering
  \begin{tabu}{@{}lll@{}}
  \toprule
  Design Option
  & Categorization
  & Examples from Related Work\\
  
  \midrule
  \begin{minipage}{12ex} \centering Modify \par Artifact\end{minipage}
  & \begin{minipage}{32ex} node (Figure~\ref{fig:spatialreferencedesign}), \\
 highlight (Figures~\ref{fig:temporalreferencedesign} and~\ref{fig:temporalreferencedesignindividual})
  \end{minipage}
  & \begin{minipage}{6.25cm}
  interaction availability filter \citep{casarin2018uau}, \\
  ``ghost'' avatar \citep{lacoche2017caf}
  \end{minipage}\\
  
  \midrule
  \begin{minipage}{12ex} \centering Add \par Artifact \end{minipage}
  & \begin{minipage}{32ex} pillar (Figure~\ref{fig:spatialreferencedesign}), \\
  pointer (Figures~\ref{fig:temporalreferencedesign}, \ref{fig:temporalreferencedesignindividual} and~\ref{fig:temporalreferencedesignindicators}), \\
  symbol (Figures~\ref{fig:temporalreferencedesign}, \ref{fig:temporalreferencedesignindividual} and~\ref{fig:temporalreferencedesignindicators_symbol})
  \end{minipage}
  & \begin{minipage}{6.25cm} 
  extended grid \citep{lacoche2017caf}, \\
  light beam \citep{peter2018vrg}, \\
  virtual drone \citep{peter2018vrg}, \\
  virtual hand \citep{sugiura2018aac}, \\
  marker \citep{welsford2020aib}
  \end{minipage}\\
  
  \midrule
  \begin{minipage}{12ex} \centering Modify \par Environment\end{minipage}
  & \begin{minipage}{32ex} location (Figure~\ref{fig:spatialreferencedesign}), \\ 
outline (Figures~\ref{fig:temporalreferencedesign} and~\ref{fig:temporalreferencedesignindividual})
  \end{minipage}
  & \begin{minipage}{6.25cm} 
  safe navigation \\ floor \citep{lacoche2017caf}, \\ 
  outline \citep{peter2018vrg}
  \end{minipage}\\
  
  \bottomrule
  \end{tabu}
\end{table}

\paragraph{Modify Artifact (MA)}
The MA design option follows the concept of temporary \textit{modifying the visual appearance of the referred artifact}, aiming to distinguish it from all others accordingly.
It is important that such a modification allows the user to detect the referred artifact, even if the process of the actual modification was not observed, i.e., the transition from normal to referred visual appearance.
We believe that this is particularly important within the context of VR, as it can not be guaranteed that the referred artifact is within the immersed user's field of view at all times, thus the change in the visual state might not be observed.
Depending on the complexity of the implemented modification to the (existing) artifact, we consider this option to be comparatively friendly in regard to the required computational resources, as it is likely that no new artifacts and geometry need to be added to the scene.
At the same time, the visual alteration of the referred artifact should be carefully considered, as potential visual mapping and data encoding may be lost through the modification of its original appearance.

\paragraph{Add Artifact (AA)}
The AA design option follows the concept of temporarily \textit{adding a visual artifact in close proximity to the referred artifact}, serving as a visual annotation.
Such an added artifact should strive to
(1) enable clear identification of the associated artifact it is intended to signal to, allowing the VR user to effectively focus on the referred artifact,
(2) be easily detectable and distinguishable from all other artifacts in the scene but at the same time
(3) not obstruct or occlude other important information in the scene.
Using this option, the visual appearance and integrity of the referred artifact are maintained in its original state, thus not losing any potentially applied visual mapping and data encoding, which are likely to be particularly relevant within analytical use cases.
Adding an artifact to the scene, although temporary, requires some practical considerations.
For instance, depending on the added artifact's complexity, such as its geometry, its addition to the scene will likely demand additional computational resources.
Thus, it is important to ensure that this is implemented in a way as to avoid a noticeable impact on the immersive application's performance.
Furthermore, based on the designer's and developer's assessments, temporarily adding (and removing) artifacts should be possible implementation-wise in a comparatively effortless and reasonable way.

\paragraph{Modify Environment (ME)}
The ME design option follows the concept of temporarily \textit{modifying existing artifacts of the environment} that are in close proximity or can otherwise be directly associated with the referred artifact.
As opposed to the AA option, rather than introducing an additional artifact to the scene, the ME option builds upon the utilization of existing elements or features in the computer-generated environment to establish a visual reference.
Naturally, this requires that the overall visualization and virtual scene are complex enough to provide such modification opportunities, to begin with.
If this requirement is met and the environment indeed provides such artifacts or alike that also allow for a semantic inference to the referred artifact, their appearance may be modified in order to act as a visual signal accordingly.
Generally, we believe this approach is likely to have a comparatively low computational impact as it utilizes existing artifacts and geometry that already exist.
At the same time, the original visual integrity of the referred artifact is maintained (as opposed to the MA design option).

\subsection{Reference Design Preface and Task Types}\label{sec:refdesignpreface}

The overall setup of the applied immersive data visualization is presented in the introduction of Section~\ref{sec:spatiotemporalreferences}.
Naturally, the reference design as collaborative information cues is inherently dependent on the applied visualization, its purpose, and its use case, e.g., what type of data is displayed using what technique.
In other words, what is the complexity and composition of the virtual 3D space?
This is important to determine what to potentially signal to, allowing for the application of the presented reference design option accordingly.
Within the presented scenario of spatio-temporal data exploration, we follow the \textit{task type} definitions as described by \citet[Chapter~3]{andrienko2006eao}, differentiating between \textit{elementary} and \textit{synoptic} tasks.
Elementary tasks are concerned with the reference to one individual data entity, e.g., one location (spatial) or one point in time (temporal).
Synoptic tasks are concerned with the reference to multiple data entities, e.g., a group of locations (spatial) or a range of multiple points in a time series (temporal).

Under consideration of the 3D Radar Chart approach and the different task types, our overall VE (see \autoref{fig:overviewimmersivedatavis}) can be described as follows.
Individual 3D Radar Charts can be uniquely identified by their geospatial location (country).
Thus, in regard to the spatial reference design, signaling both to an individual as well as multiple locations should be possible.
Each 3D Radar Chart features multiple time-series data variables.
It should be possible to refer to a single point as well as to a range of consecutive points in time.
Such time point and time range references should be possible both across all time-series data variables as well as just for individual ones.

Under consideration of this scene understanding, we can set out to design and implement spatial and temporal references as collaborative information cues as described throughout the remainder of this section.
Additionally, we provide a supplemental 360\degree~interactive web application that can be viewed online, illustrating all reference designs from a VR user's POV.\footnote{Supplemental 360\degree~interactive web application illustrating all reference designs as described throughout Sections~\ref{sec:spatialreferencedesign} and~\ref{sec:temporalreferencedesign}: \textcolor{cyan}{  \href{https://vrxar.lnu.se/apps/tdrc-ref-360/}{vrxar.lnu.se/apps/tdrc-ref-360/}} \label{foot:360}}

\subsection{Spatial Reference Design}\label{sec:spatialreferencedesign}

\begin{figure}
    \centering
    \includegraphics[width=.9\columnwidth]{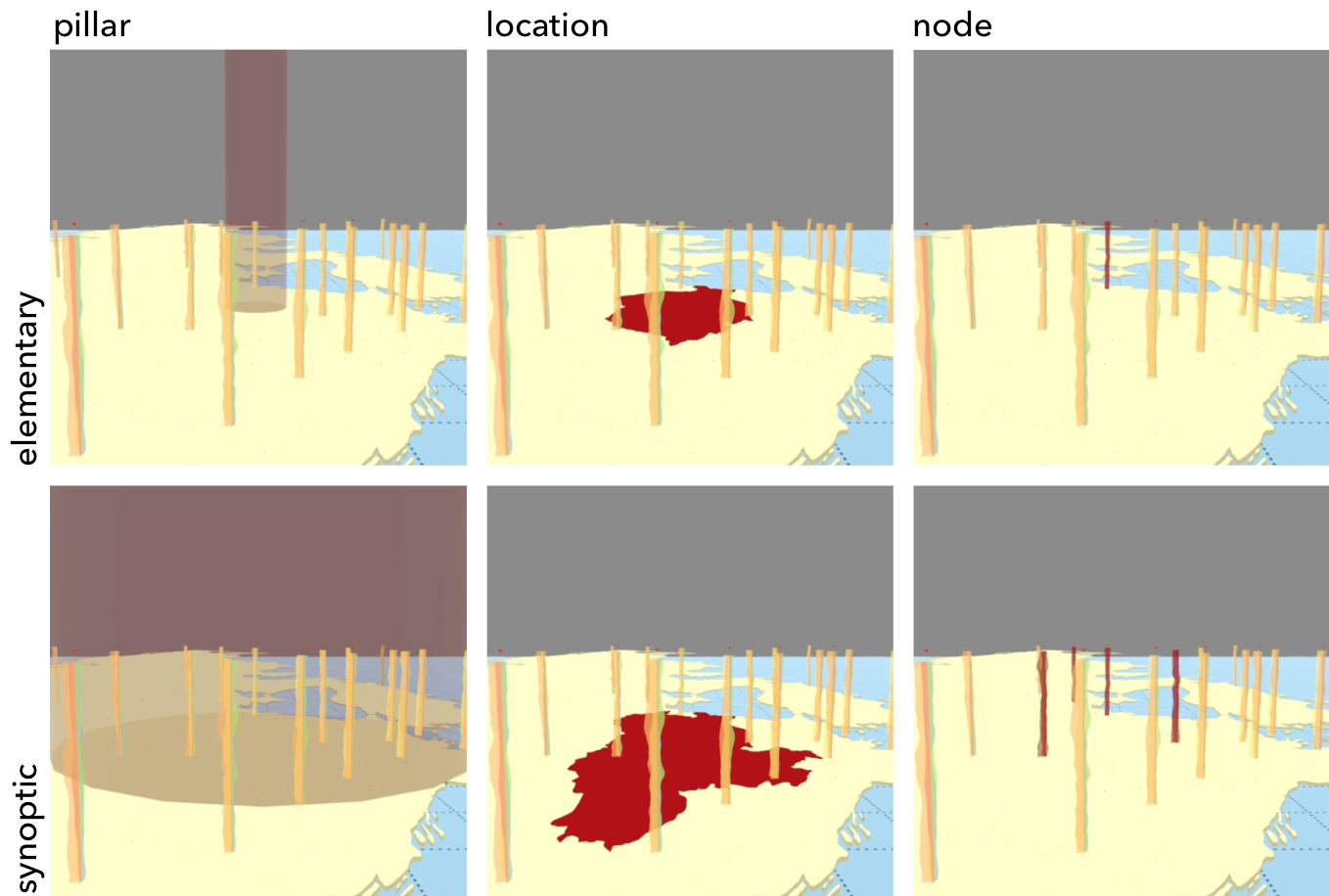}
    \caption{Spatial reference design approaches from the VR user's POV: \textit{pillar}, \textit{location}, and \textit{node}.
    Elementary task reference at the top, and synoptic at the bottom row.
    Each of the presented task reference configurations refers to the same location(s).}
    \label{fig:spatialreferencedesign}
\end{figure}

For the purpose of referring to specific 3D Radar Chart instances in the VE, we designed three different spatial reference approaches in accordance with the presented design options (see \autoref{fig:spatialreferencedesign}).
First, the \textit{pillar} design follows the AA option, creating a semi-transparent cylinder with the 3D Radar Chart at its center, surrounding it accordingly.
The pillar's height is scaled to make it appear to ``shine from the top down'' in the VE, similar to a spotlight.
Second, the \textit{location} design follows the ME option, modifying the color of the extruded country polygon on the floor that each 3D Radar Chart is directly associated with.
Third, the \textit{node} design follows the MA option, visually separating the referred 3D Radar Chart from all others by uniquely coloring all its data variable axes.

\subsection{Temporal Reference Design}\label{sec:temporalreferencedesign}

\begin{figure}
    \centering
    \includegraphics[width=.9\columnwidth]{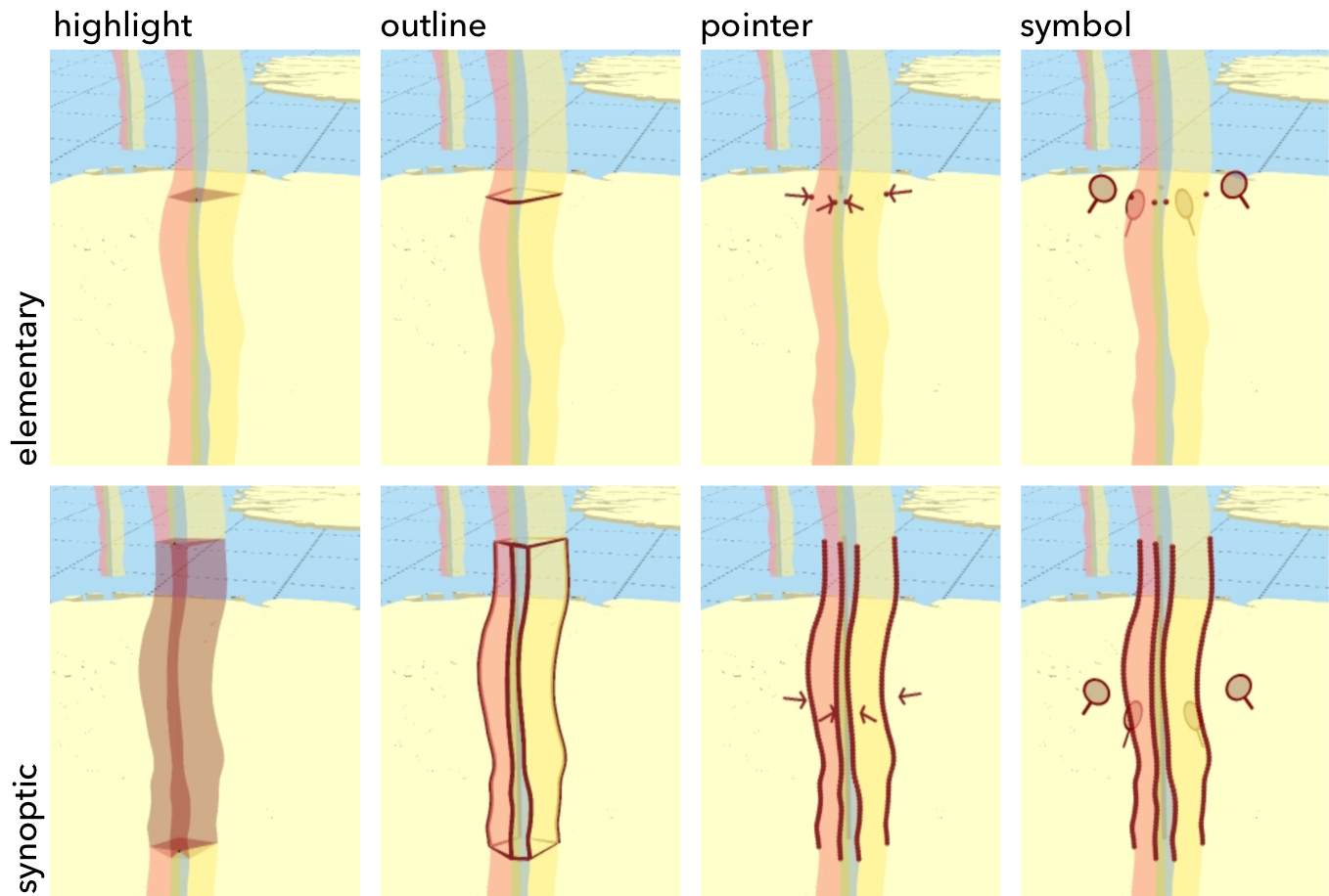}
    \caption{Temporal reference design approaches from the VR user's POV: \textit{highlight}, \textit{outline}, \textit{pointer}, and \textit{symbol}.
    Elementary task reference across all data variables at the top, and synoptic at the bottom row.
    Each of the presented task reference configurations refers to the same time point/range.}
    \label{fig:temporalreferencedesign}
\end{figure}

For the purpose of referring to single points in time as well as time ranges (multiple consecutive points in a time series) within a 3D Radar Chart instance, we designed four different temporal reference approaches in accordance with the presented design options (see \autoref{fig:temporalreferencedesign}).
First, the \textit{highlight} design follows the MA option, visualizing a colored mesh for the time point across all data variables as elementary task reference, respectively coloring the time range segments in each data variable axis as synoptic task reference.
Second, the \textit{outline} design follows the ME option, creating a closed visual loop along the outside of all included temporal data points.
Third, the \textit{pointer} design follows the AA option and adds two artifacts to the visualization as reference: Each included temporal data point is encapsulated by a small visual sphere, further assisted through a juxtaposed 3D pointer model that directly indicates the respective time point or time range.
Fourth, we implemented a \textit{symbol} design, also based on the AA option and following a similar approach as the pointer design.
However, instead of a pointer, we decided to juxtapose the virtual sphere with a symbol that can be interpreted by the user to infer further meaning.
As a practical illustration, we decided to use a magnifying glass symbol in a ``\textit{let us investigate this} [temporal reference]'' analogy.

\begin{figure}
    \centering
    \includegraphics[width=.9\columnwidth]{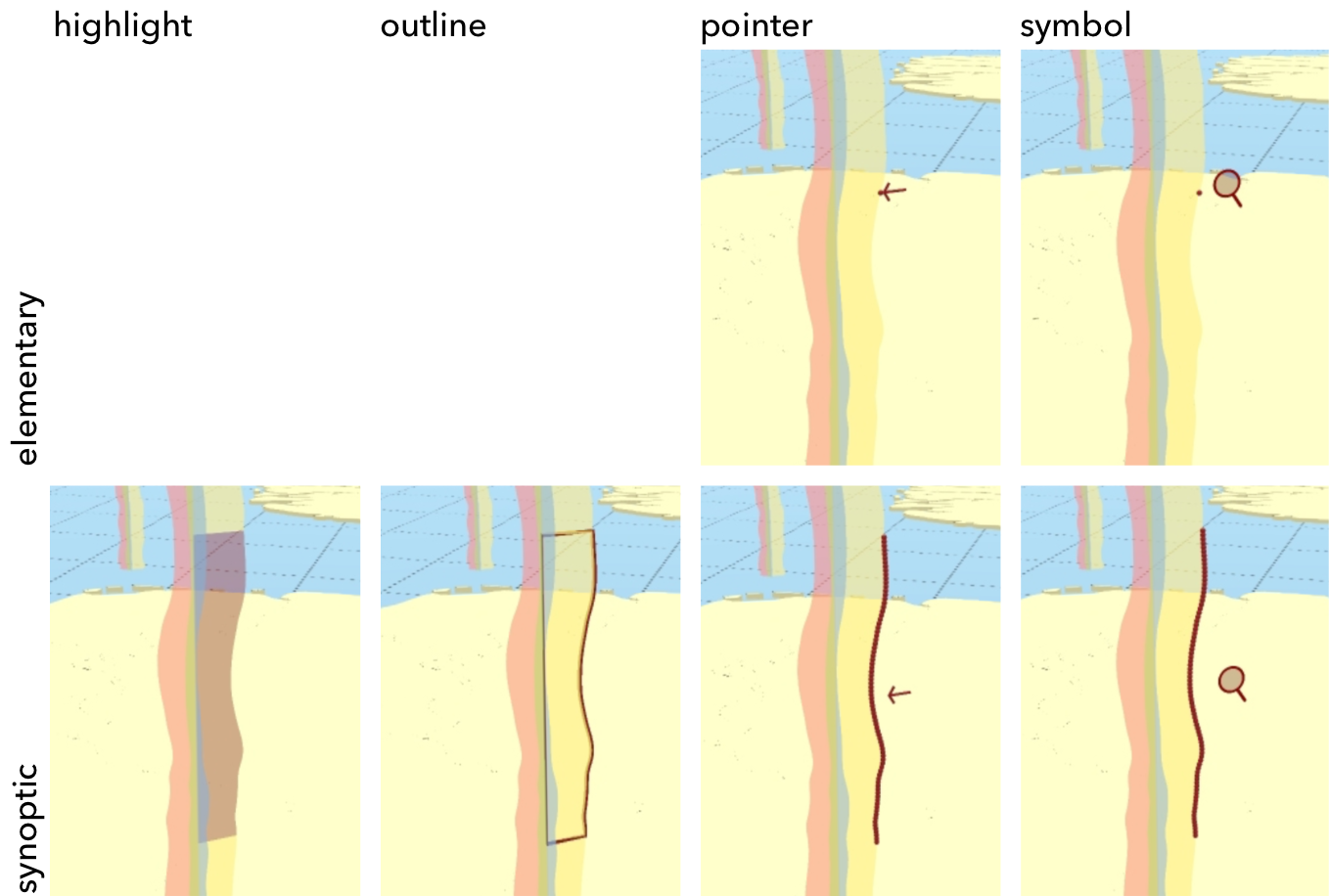}
    \caption{Temporal reference design approaches from the VR user's POV for individual data variables: \textit{highlight}, \textit{outline}, \textit{pointer}, and \textit{symbol}.
    Elementary task reference for individual data variables at the top, and synoptic at the bottom row.
    Each of the presented task reference configurations refers to the same time point/range.
    Note: Due to the nature of the \textit{highlight} and \textit{outline} approaches with respect to the 3D Radar Chart visualization, implementations for an elementary task reference in an individual data variable axis are not feasible.}
    \label{fig:temporalreferencedesignindividual}
\end{figure}

\begin{figure}
    \centering
    \includegraphics[width=.9\columnwidth]{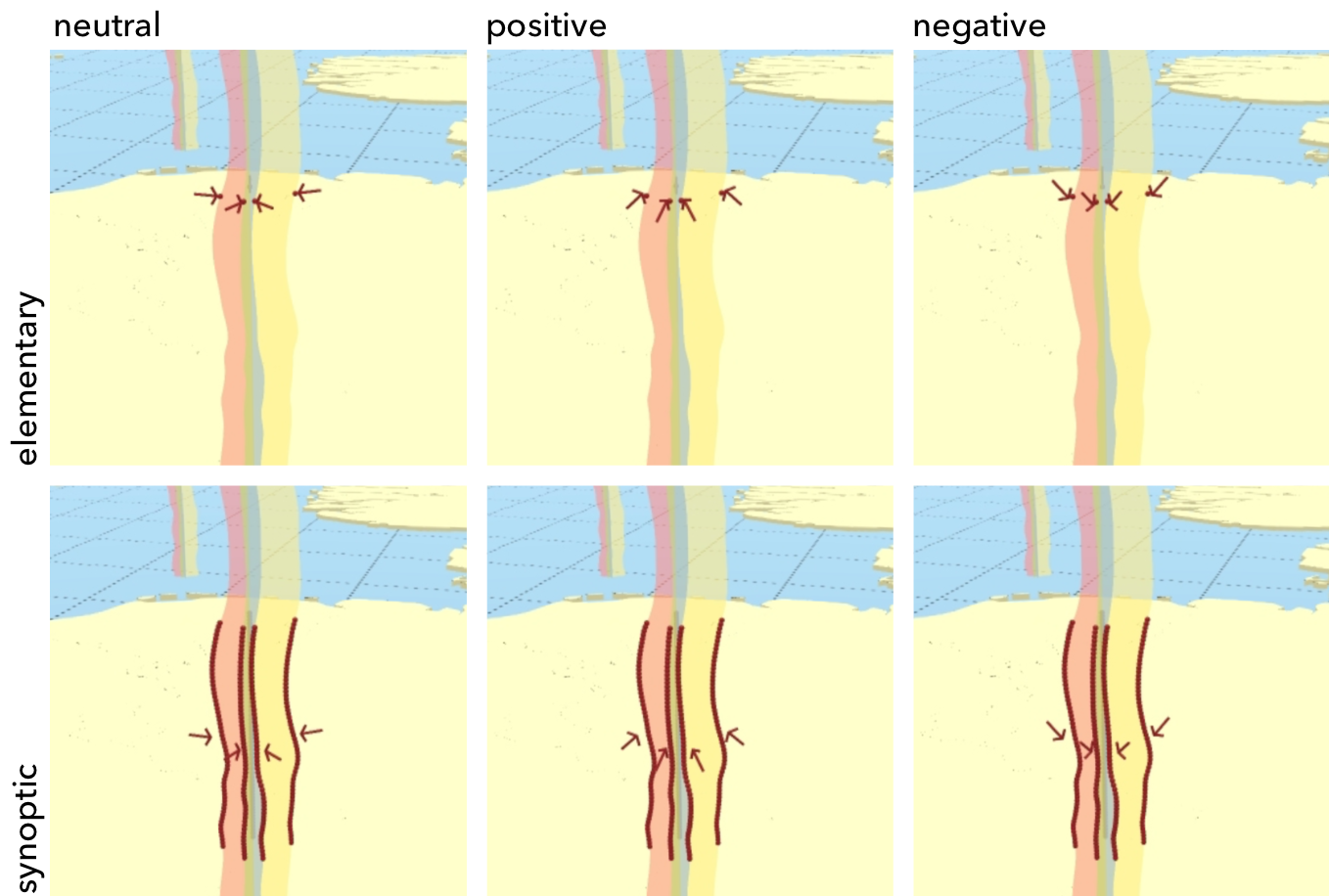}
    \caption{Temporal reference indicator design approaches from the VR user's POV using \textit{pointer} in a \textit{neutral} (pointing straight), \textit{positive} (pointing upwards), or \textit{negative} (pointing downwards) manner.
    Elementary task reference for all data variables at the top, and synoptic at the bottom row.
    Each of the presented task reference configurations refers to the same time point/range.
    }
    \label{fig:temporalreferencedesignindicators}
\end{figure}

\begin{figure}
    \centering
    \includegraphics[width=.9\columnwidth]{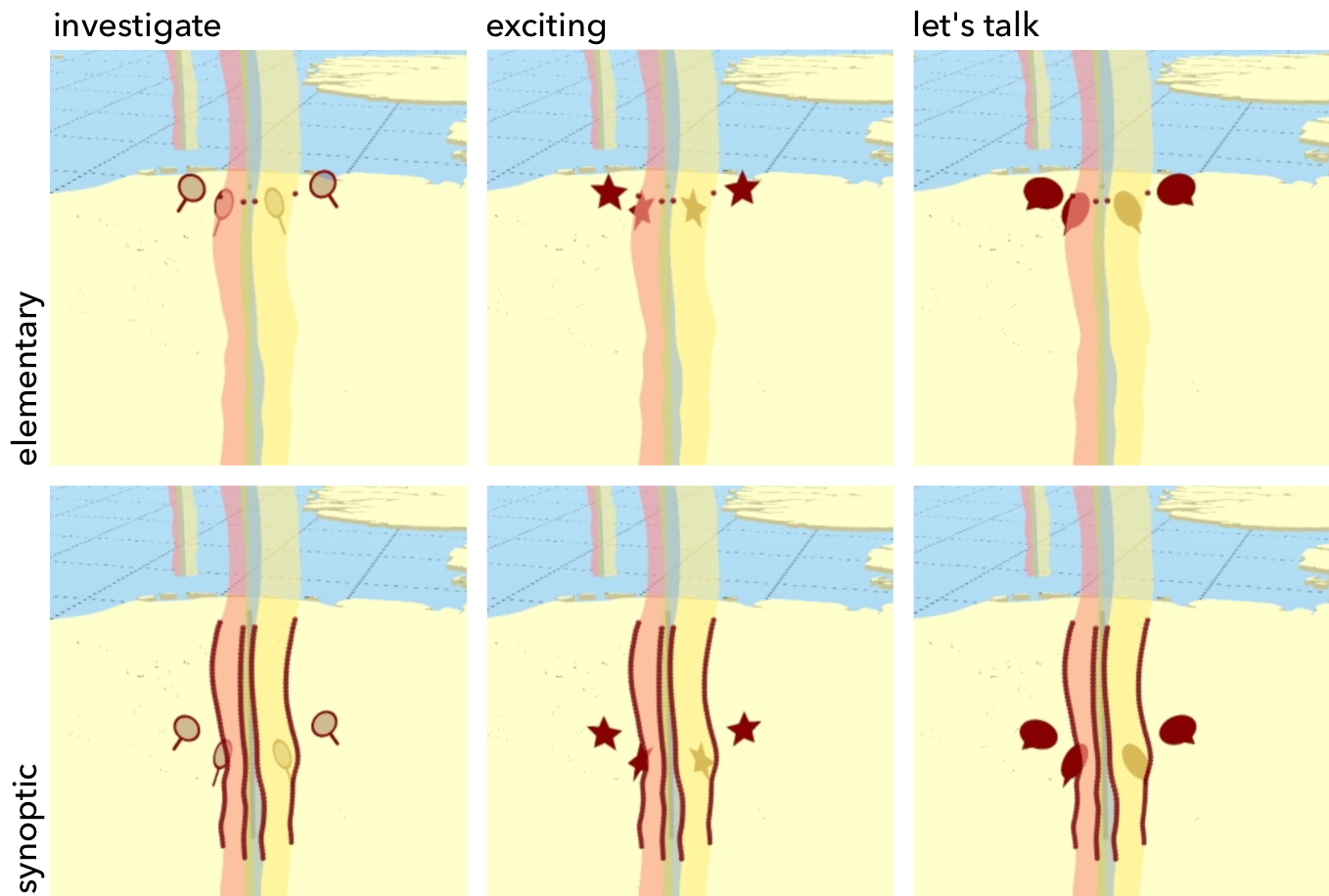}
    \caption{Temporal reference indicator design approaches from the VR user's POV using \textit{pointer} in an \textit{investigate} (magnifying glass), \textit{exciting} (star), or \textit{let's talk} (speech bubble) styling.
    Elementary task reference for all data variables at the top, and synoptic at the bottom row.
    Each of the presented task reference configurations refers to the same time point/range.
    }
    \label{fig:temporalreferencedesignindicators_symbol}
\end{figure}

The complexity of the 3D Radar Chart visualization allows for different temporal reference configurations.
Making a reference across all data variables is illustrated in \autoref{fig:temporalreferencedesign}, whereas references in individual data variables are presented in \autoref{fig:temporalreferencedesignindividual}.
Additionally, we are interested to explore different configurations of the pointer and symbol approaches, as presented in \autoref{fig:temporalreferencedesignindicators} and \autoref{fig:temporalreferencedesignindicators_symbol} respectively.
For instance, the placement of the pointer indicator could encode further analysis-related information, such as through a \textit{neutral}, \textit{positive}, or \textit{negative} pointing direction, e.g., to provide a comparison to a prior data value or to indicate an overall trend across the referred time range.
Similarly, the application of different symbols could indicate an additional collaborative information cue, for instance, to provide a reason why the collaborator is making a reference in the first place, such as to \textit{investigate} because they found something they deem \textit{exciting}, or because they want to further \textit{talk} about the referred data.

\subsection{Technologies and Implementation}\label{sec:techandimplementation}

The immersive spatio-temporal data visualization using 3D Radar Charts is developed using Unity 2019.3 and the SteamVR Plugin for the Unity package.
An HTC Vive HMD (1080x1200 pixel resolution per eye, 90 Hz refresh rate) is utilized for visual immersion, allowing the VR user to naturally look around and make observations.
Within the scope of our investigation, no additional interactive features are provided.
All presented spatio-temporal reference design approaches (see Section~\ref{sec:spatialreferencedesign} and Section~\ref{sec:temporalreferencedesign}) were implemented using the base functionalities as provided in Unity.
For the purpose of the user study, to make references on demand in the VR environment, we developed a complementary web application in HTML5, CSS, and JavaScript.
A WebSocket Secure server based on Node.js functions as the network communication interface between Unity and the web application.
A template implementation that illustrates this workflow is available online.\footnote{GitHub repository of the \emph{Unity - Connect via WebSocket server to JavaScript client} project: \href{https://github.com/nicoversity/unity_wss_js}{github.com/nicoversity/unity\_wss\_js}
}
This approach has several advantages:
\begin{enumerate}
    \item We are able to trigger the various reference configurations in the VE, thus simulating potential input from a real-world collaborator.
    
    \item We can conveniently prepare presets of reference configurations in anticipation of their systematic evaluation (see Section~\ref{sec:methodology}).
    
    \item The implementation provides a modular and easily extendable example of an application programming interface (API) to make references in the VR environment from the outside, allowing integration with other tools and applications in the future.
\end{enumerate}

\section{Evaluation Methodology}\label{sec:methodology}

In order to assess the designed spatio-temporal reference approaches, we conducted an empirical evaluation.
This section provides details about the overall study design as well as the applied measures.

\subsection{Physical Study Space and Virtual Environment}

The study was planned as one-on-one sessions involving only the participating user and the researcher.
The researcher was responsible for the practical conduction of the study sessions, i.e., taking care of the study moderation, ensuring all involved hard- and software were functioning as intended, and collecting the data.
Each study session was conducted in our research group lab, providing a two-by-two meter area for the VR user, a desk as the researcher's workstation, as well as a desk for the participating user that is physically partitioned from the researcher's workspace.
The research group lab is big enough for the researcher and participant to conduct the study comfortably.
The researcher remained at their workstation to moderate the study, operate the presentation of the various spatio-temporal reference design approaches using the implemented web interface, and take notes in regard to the data collection.
The participant was first briefly seated at their desk to complete an informed user consent form and was then located exclusively within the designated VR user area for the remainder of the study session.

We set up the VE as presented in \autoref{fig:overviewimmersivedatavis}, placing a total of 39 3D Radar Charts across different European countries.
Each 3D Radar Chart featured five data variable axes, each comprised of a time series with 150 events.
The VR user was placed in Central Europe, able to move freely within the calibrated two-by-two-meter area and make observations.
There was one 3D Radar Chart placed within the VR user's physical movement area, all others were beyond their reach.
No additional interactive features were provided.

More specifically, each 3D Radar Chart in the VE featured a total height, i.e., vertical length, corresponding to $1.0~meters$.
All 3D Radar Charts were placed to hover $0.4~meters$ above the virtual floor, thus reaching an effective height of $1.4~meters$ in the VE.
The 3D Radar Chart used to display all temporal reference configurations was placed directly at the center of the VR user's calibrated two-by-two meter area, enabling them to move freely around to investigate that 3D Radar Chart from all sides if so desired.
The distance between the center of the VR user's area and the spatial reference for the elementary task was approximately $8.48~meters$ in the virtual space.
Furthermore, the distances between the center of the VR user's area and the spatial references for the synoptic tasks were approximately $5.73$, $6.45$, $8.48$, and $10.24~meters$.
These distances resulted from the properties of the underlying dataset used within the study (see Section~\ref{sec:spatiotemporalreferences}).
For additional impressions, please refer to the supplemental 360\degree~interactive web application as provided in Section~\ref{sec:refdesignpreface} (Footnote \ref{foot:360}), visually illustrating all reference configurations as used in the study.

\subsection{Measures}\label{sec:measures}

To systematically evaluate the different spatio-temporal reference design approaches, we decided to apply subjective methods and center our data collection around three measures, i.e., \textit{aesthetics}, \textit{legibility}, and \textit{general user preference}.
Within the scope of our investigation, we define \textit{aesthetics} as how much one appreciates the visual design and appeal of a reference design approach, and whether or not one finds it pleasing and beautiful to look at.
We define \textit{legibility} as how well one can understand, determine, and detect what is highlighted, and how clear it is what to focus on.
Collecting quantitative ratings for these two metrics for each reference design allows for respective comparisons.
Additionally, the third metric is concerned with the \textit{user's general preference} based on pair-wise comparisons, indicating which one of two reference design approaches they would rather work with if they were to use such an application frequently.
This allows for respective tallying of the results,\footnote{The pair-wise preference comparison within the scope of our investigation was inspired by prior experiences of applying the NASA Task Load Index \citep{hart2006ntl}, particularly its \textit{weighting} process.} providing an overall preference indication as well as functioning as a potential tie-breaker between two approaches in the case of equal aesthetics and legibility ratings.

\subsection{Task}

To assess aesthetics, legibility, and general user preference for the different reference design approaches, we prepared presets for all the different configurations, and tasked the participants to rate them following the Thinking Aloud protocol.
In particular, we inquired assessments for all the elementary and synoptic spatio-temporal references as presented throughout \autoref{fig:spatialreferencedesign}, \autoref{fig:temporalreferencedesign}, and \autoref{fig:temporalreferencedesignindividual}.
Additionally, we also inquired about the user's assessment for the temporal reference design of the pointer and symbol approaches in general (one combined assessment each for pointer and symbol) as presented in \autoref{fig:temporalreferencedesignindicators} and \autoref{fig:temporalreferencedesignindicators_symbol}.
The reference presets were generally configured to make the same reference, e.g., to refer to the same location (spatial) or time range (temporal).
The assessments from each participant were collected in random order throughout two stages.
First, they were presented with the individual reference presets and tasked to \textit{rate} their perceived aesthetics and legibility on a 7-point Likert scale.
They were asked:
\begin{itemize}
\setlength\itemsep{0em}
    \item Aesthetics: On a scale from 1 (aesthetically unpleasing) to 7 (aesthetically pleasing), how would you rate this approach of [~spatial / temporal~] referencing?
    \item Legibility: On a scale from 1 (not at all) to 7 (very well), how clearly can you determine what is [~spatially / temporally~] referenced?
\end{itemize}
Second, once they provided numerical assessments for all the presented references, we inquired about their general preference for one over the other in a series of pairwise comparisons.
For each logical category, i.e., spatial elementary, spatial synoptic, temporal elementary, and so on, we created all possible pair permutations and asked:
\begin{itemize}
\setlength\itemsep{0em}
    \item Out of these two [~spatial / temporal~] referencing approaches, which one do you prefer?
\end{itemize}

\subsection{Study Procedure}

Each study session followed the same procedure of three stages:
(1)~introduction,
(2)~task: aesthetics and legibility, and
(3)~task: pair-wise preference comparison.
Each session was aimed to take approximately 30 to 40 minutes (10~min introduction, 20--30 min immersed in VR).
Each participant filled out an informed user consent, after which some demographic information (professional background, prior VR experience) was collected.
The researcher presented the overall context and scenario of the immersive application in regard to its analytical and collaborative aspects, ensuring that each participant understood the purpose and composition of the VE.
Participants were then provided with a brief warm-up allowing them to familiarize themselves with wearing the HMD, and with the VE.
Once they felt comfortable, they proceeded to the tasks.
Based on random order, the participant's aesthetics and legibility ratings for the different spatio-temporal reference approaches were inquired via the Thinking Aloud protocol.
Afterward, they were asked to state their general preference for one approach over the other.
Based on random order, each participant provided distinct answers for each pairwise comparison.
The researcher noted all ratings and preferences on a predefined task answer sheet.
Furthermore, throughout both task stages, the participants were allowed to provide additional remarks as desired that were also noted by the researcher.
Finally, they were thanked for their participation in the study and sent off.

\subsection{Ethical Considerations}

We followed general ethical considerations for the work with human participants within the scope of human-computer interaction research \citep{nent2016gfr,swedishresearchcouncil2017grp}.
The presented empirical evaluation was conducted between April and June 2021 during the, at the time ongoing, global COVID-19 pandemic.
Consequently, additional practical precautions were implemented.
We closely monitored and followed all national, regional, and local health and safety recommendations according to the respective authorities.
Study sessions were only conducted when all parties (participant and researcher) were symptom-free, and also kept recommended physical distance at all times.
The researcher was wearing a face mask at all times.
Face masks and hand disinfection gel were freely available to each participant.
All involved technical equipment was carefully sanitized between study sessions.

\section{Results}\label{sec:results}

\begin{figure}
    \centering
    \includegraphics[width=.95\columnwidth]{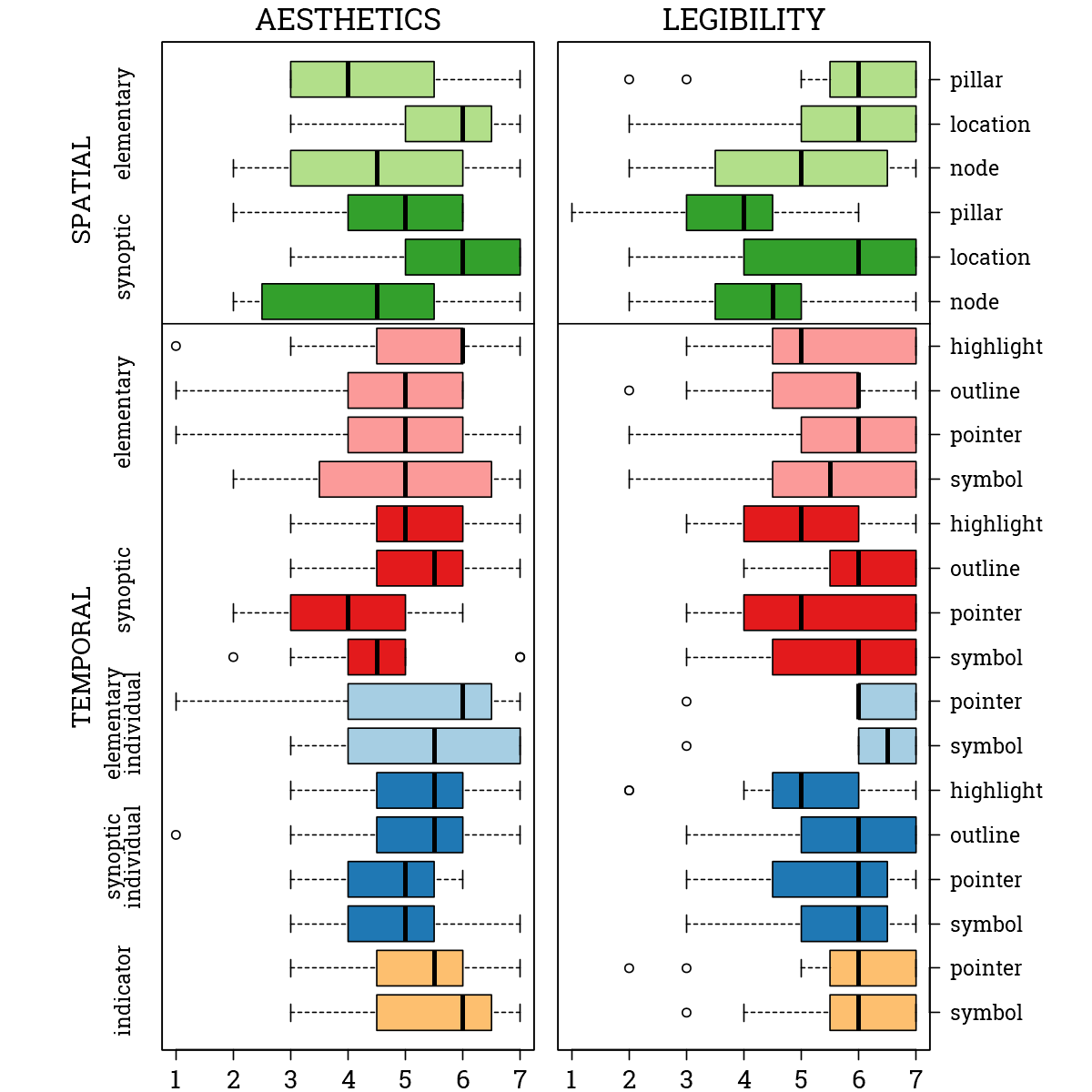}
    \caption{The rating scores ($n=12$) of the implemented reference design approaches, in terms of \textit{aesthetics} (left) and \textit{legibility} (right).} 
    \label{fig:aesthetics_legibility}
\end{figure}

\begin{figure}
    \centering
    \includegraphics[width=.9\columnwidth]{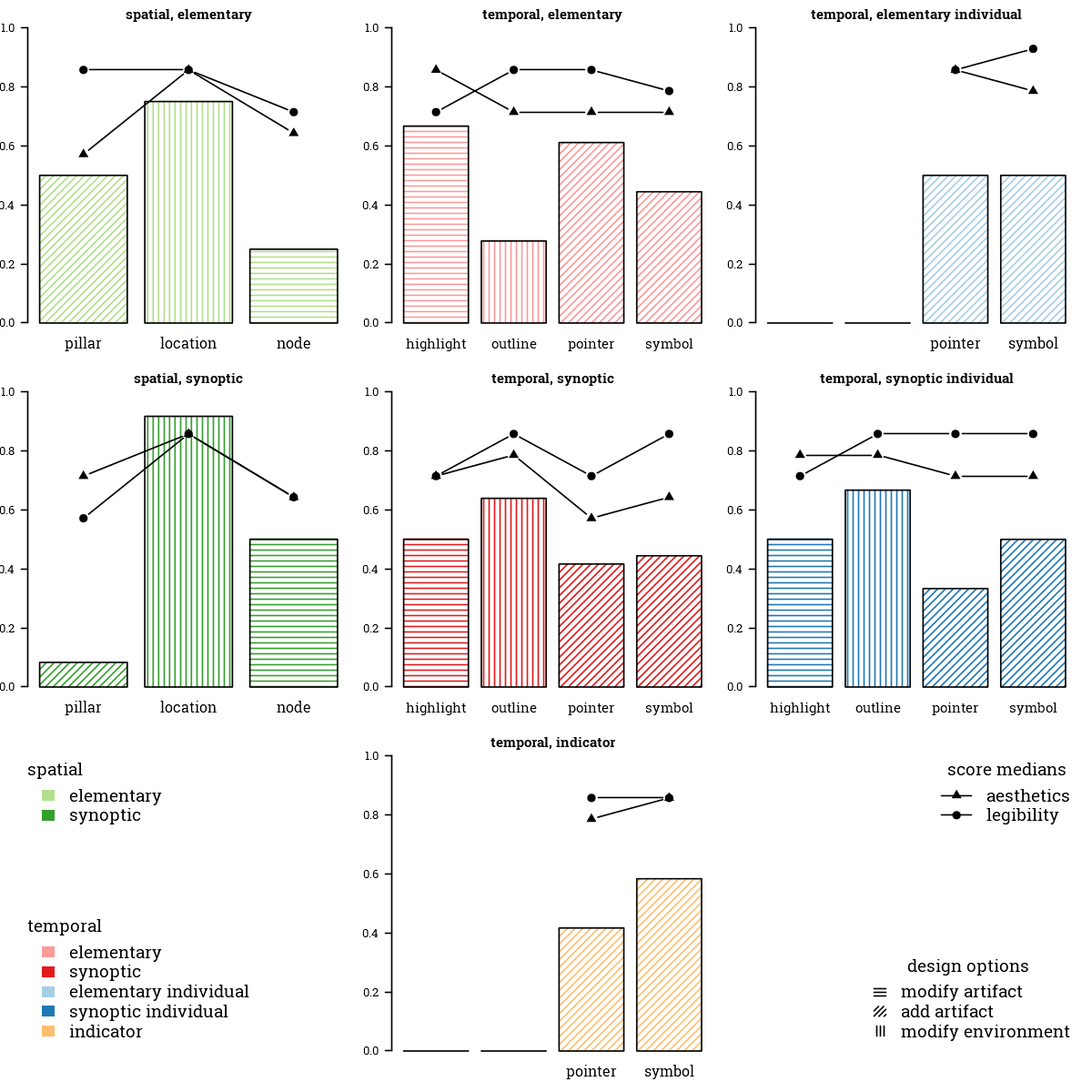}
    \caption{Bar plots -- the participants' preferences for each of the seven categories (\textit{spatial, elementary~/ synoptic}; \textit{temporal, elementary~/ elementary individual~/ synoptic~/ synoptic individual~/ indicator}), when asked to select between the different available approaches (pair-wise comparison).
    Given the unequal number of options in each category, the values are here presented normalized.
    Line plots -- the (normalized) medians for the \textit{aesthetics} and \textit{legibility} ratings (see \autoref{fig:aesthetics_legibility}).} 
    \label{fig:prefs}
\end{figure}

We recruited $n = 12$ participants from a mixture of different academic backgrounds (5~Computer and Information Science, 5~Linguistics and Language Studies, 2~Forestry and Wood Technology).
Eight participants reported having just \textit{a little} prior experience with VR, three \textit{average}, and one \textit{a lot}.
None of the participants reported any visual perception issues based on the applied color coding in the VR environment.
\autoref{fig:aesthetics_legibility} presents the results of the participants' ratings for the aesthetics and legibility measures.
The results of the pair-wise preference comparison (combined with the rating medians) are presented in \autoref{fig:prefs}.

Some participants provided additional remarks for the different reference design approaches.
For instance, participants stated that they can envision usefulness and relevance for both pointer and symbol temporal reference design approaches (see \autoref{fig:temporalreferencedesignindicators} and \autoref{fig:temporalreferencedesignindicators_symbol}) in a real-world scenario.
According to them, the pointer approach subjectively represented more precision and urgency, while the symbol one was easier to recognize and featured better clarity.
One participant stated that the pointer approach makes more sense during synchronous collaboration (as the users are likely to talk to each other), while the symbol approach may be better suited for asynchronous collaboration (encoding an additional semantic meaning).
It was also noted that a visual connection to the 3D Radar Chart's origin (time axis) is missing and would be preferred in the synoptic reference, as included in the outline approach for an individual dimension reference (see \autoref{fig:temporalreferencedesignindividual}).
Some participants were unsure whether the pointer for the synoptic referencing task was referring to the entire time series or to one specific time event.
One participant expected the spatial pillar approach to be multiple small ones as opposed to one large one when making a synoptic reference (see pillar in \autoref{fig:spatialreferencedesign}).
Another one stated that it might increase the legibility of the spatial location approach by additionally slightly extruding the respective country model (see location in \autoref{fig:spatialreferencedesign}).
As in the investigation reported by \citet{peter2018vrg}, participants made suggestions for some hybrid designs, combining two approaches into one, such as spatial elementary pillar+node, spatial elementary location+node, temporal elementary highlight+pointer, temporal synoptic outline+symbol without the visual spheres, and temporal synoptic pointer but without the pointer using just the visual spheres -- similar to the marker approach by \citet{welsford2020aib}. 

\section{Discussion}\label{sec:discussion}

\subsection{Spatial References}\label{sec:discussion:spatialref}

Out of the three implemented spatial reference designs, the \textit{location} approach was most favored by the participants universally across both elementary and synoptic task configurations.
Perceived aesthetics and legibility ratings were comparatively higher than for the \textit{node} and \textit{pillar} approaches.
The realization of this approach was possible in our use case due to the availability of individual extruded country polygons.
Using the features of the VE seemed to have been appreciated by the participants, allowing the identification of either one or multiple referred artifacts, while at the same time maintaining the original visual composition of the data visualizations.
Within the presented CIA context, this was very much favored by the participants.
Interestingly, the user preference in favor of the implemented \textit{location} approach is somewhat in contrast to the results reported by \citet{lacoche2017caf}, where their safe navigation floor approach rated worse compared to the others.
The actual approach was visually similar in both cases, but the signal's intent differed: While our purpose was to actively guide the user to the highlighted area, theirs was instead to avoid it \citep{lacoche2017caf}.

Comparing the \textit{node} and \textit{pillar} approaches, it is interesting to see the disparity in regard to their assessments across the elementary and synoptic task configuration.
The \textit{pillar} approach was rated better for referring to one artifact, likely because it allowed the participants to quickly and easily identify the referred artifact (legibility), while it took them subjectively a bit longer to identify the \textit{node} design.
On the other hand, the synoptic approach of using one large pillar to highlight a group of artifacts did not scale accordingly, as it was not clear to them what artifacts are referred to exactly.
Maybe such an approach would be better suited if one was to make a reference to an otherwise not specified large spatial area, instead of a group of identifiable artifacts.
These results are quite interesting, as they indicate favor for different design approaches based on the task configuration.
Arguably from a user interface perspective, one would often strive to apply a uniform design strategy, implementing the same approaches for the same, or similar, tasks.
Within the context of the presented scenario, a reasonable design choice follows: Should one implement the same design approach across both tasks, thus following a rather coherent design, or instead implement different approaches, each in better support for their given task?
Clearly, this requires careful consideration from the application's interface designer, weighing the pros and cons of either approach given the purpose and task at hand.
We believe that in the presented data analysis context, legibility and user preferences independent of the design approach are more important, allowing analysts to be as precise as possible in their referencing and subsequent collaboration.
Consequently, if environmental features were not available for implementation of the \textit{location} approach, we would instead apply a mixture of \textit{pillar} for elementary referencing and \textit{node} for synoptic.
The \textit{pillar} approach is conceptually similar to the light beam technique as presented by \citet{peter2018vrg}, who reported mixed results from their evaluation, describing that sometimes the participants would consider the light beam being part of the VE instead of a being a dedicated signal.
The realistic design of the light beam seemed to have conflicted with the realistic setting of their VE, which prevented participants from clearly identifying the signal through the VR-Guide \citep{peter2018vrg}.

\subsection{Temporal References}\label{sec:discussion:temporalref}

No distinct temporal reference design approach was favored across elementary and synoptic tasks within the 3D Radar Chart scenario.
For point-in-time (elementary) referencing, the participants generally preferred the \textit{highlight} approach.
Interestingly, while it scored better aesthetics ratings, its median for legibility is the worst compared to the three other approaches.
The \textit{pointer} approach is a close runner-up in regard to user preference, scoring generally better legibility ratings.
It seems that the participants liked the analogy of literally ``pointing to a point in time'' -- even though we used a rather abstract approach instead of a realistic one \citep{sugiura2018aac}.

Similar to the results presented by \citet{peter2018vrg}, we identified trends towards the participants' favor of the \textit{outline} approach compared to all others for the synoptic task configuration, both across all data dimensions and for the individual one.
This is particularly interesting and also somewhat odd, as the \textit{outline} approach received the lowest user preference for the elementary task, even though it scored comparatively good in regard to aesthetics and legibility ratings.
These results reveal a similar preference disparity as for the \textit{node} and \textit{pillar} approaches (see Section~\ref{sec:discussion:spatialref}), requiring careful consideration in favor of or against a uniform interface design.

We implemented two approaches based on the AA option using different indicator types: \textit{pointer} and \textit{symbol}.
The \textit{symbol} approach was rated slightly better when directly compared to the \textit{pointer} one (see different indicator configurations as illustrated in \autoref{fig:temporalreferencedesignindicators}).
However, the results indicate trends toward a rather equal preference for both approaches when examining the bigger picture.
We thought the participant's comment in regard to using the \textit{pointer} approach during synchronous collaboration, while encoding additional meaning in the \textit{symbol} indicator during asynchronous collaboration, in a more annotation-like manner, was particularly interesting, encouraging some further investigation.
No major advantages or disadvantages for one over the other were identified, making both potentially valid approaches depending on the reference task and purpose.

Within the presented context and scenario, we argue as follows.
The \textit{outline} approach appears to be a favored referencing approach for making time-series references (synoptic), especially under consideration of the comparatively good aesthetics and legibility ratings.
However, implementation as a temporal reference for an elementary task configuration of an individual data variable was not possible using the \textit{outline} approach (neither for the \textit{highlight} one).
Given the preference for the \textit{pointer} approach over the \textit{symbol} one for the elementary task, we can see its application respectively, thus again recommending a mixture of different design approaches across elementary and synoptic tasks for making temporal references.

\section{Conclusion and Future Work}\label{sec:conclusion}

We set out to investigate spatio-temporal reference design approaches within the context of CIA.
Collaborative information cues, such as pointing and referring to artifacts, are important in general, but arguably even more so in scenarios that involve the use of immersive technologies where more traditional co-located input is no longer conventionally available.
To address such shortcomings, we implemented different design approaches to make spatial and temporal references within an IA environment.
The design approaches were guided through relevant prior work in the field and the derivation of three design options for the creation of visual references.
We hope that the presented options can assist practitioners with the design of similar information cues on a general level in the future, independent of their application within the context of CIA.
Under consideration of the available features in the VE as well as based on different referencing tasks (elementary and synoptic), we empirically evaluated the implemented reference approaches in regard to aesthetics, legibility, and general user preference.
The participants preferred the presented \textit{location} approach as a spatial reference, implemented through the modification of environmental features that can be directly associated with the referred artifact.
Based on the results and discussion in regard to temporal references, it appears that different design approaches are preferable depending on the task: While the \textit{pointer} approach seems like a valuable option for referencing to single points in time, the \textit{outline} approach was clearly favored for all time-series references.
In multiple instances of the results (spatial \textit{node} versus \textit{pillar}, temporal \textit{outline} versus all others), trends indicate user preference for different approaches depending on the task (elementary versus synoptic).
This requires careful consideration in regard to the general user interface design, as the implementation of different approaches may be in conflict with an otherwise desired uniform design throughout various aspects of an application's interface.
Based on the variety of different reference task configurations, it may be valuable to explore the design of multiple approaches, each serving best its own purpose.
This can be particularly valuable within data analysis-related scenarios, referring to different types and aspects of abstract data visualization, ensuring it is clear to the collaborator what their respective partner is referring to.

In fact, informed by the outcomes of our study, we utilized the location (spatial) and symbol (temporal; magnifying glass) referencing approaches in a collaborative user study where two analysts were tasked with the investigation of a spatio-temporal dataset to make confirmative analysis assessments \citep{reski2022aee}.
The system setup was based on heterogeneous display and interaction technologies, comprising an immersive VR environment based on the 3D Radar Chart visualization approach (as presented throughout this article) as well as a non-immersive interface. 
Each interface featured various collaborative information cues to enable bidirectional spatio-temporal referencing across the different display modalities in real-time.\footnote{Due to the \cite{reski2022aee} study setup, referencing was only elementary (country) for the spatial dimension, but could be either elementary or synoptic (time event or time range) for the temporal dimension across all data variables. } 
The results of that follow-up study, particularly within the context of a more applied real-world CIA use case, enabled us to further investigate such referencing approaches.

We see the potential for further investigations.
First, the design and evaluation of hybrid reference approaches as suggested by some participants seem intriguing, combining aspects of various approaches into new ones.
Second, the application of the presented design options for the creation of reference approaches as collaborative information cues in other contexts is intriguing, allowing for further iteration and the addition of conceptual principles to the options.
Third, naturally, we intend to utilize the gained insights from the evaluation and apply such spatio-temporal reference design approaches in a practical real-world scenario, further evaluating the collaborative interplay between multiple analysts when using immersive technologies.
And fourth, while we envisioned the design approaches as presented here only as temporary references that are removed after a short while from the VE, it would be interesting to investigate similar approaches for more persistent data annotation -- a topic that is also largely unexplored within the context of CIA \citep{fonnet2021soi}.

\section*{Acknowledgments}
\noindent
The authors wish to thank all the participants of the user study.
This work was partially supported by the ELLIIT environment for strategic research in Sweden.

\section*{Author Contributions}
\noindent
All authors (NR, AA, and AK) devised the research scope.
NR and AA designed the empirical evaluation.
NR reviewed the literature, developed all technical parts of the system (immersive VR environment, complementary web application, network communication interface), recruited study participants, and conducted the empirical evaluation (user study) as well as the data collection.
AA and NR conducted the data analysis.
NR and AA wrote the manuscript.
All authors discussed and reviewed the manuscript.

\bibliography{references01.bib}

\end{document}